\documentclass[a4paper,9pt]{article}

\usepackage{INTERSPEECH2020}
\usepackage{multirow}
\usepackage{wrapfig}
\title{From Speaker Verification to Multispeaker Speech Synthesis,  \\ Deep Transfer with Feedback Constraint }
\name{Zexin Cai$^{\star}$, Chuxiong Zhang$^{\dag}$, Ming Li$^{\star, \dag}$}
\address{
  $^{\star}$Electrial \& Computer Engineering, Duke University, Durham, NC, United States  \\
$^{\dag}$ Data Science Research Center, Duke Kunshan University, Kunshan, China }
\email{ming.li369@duke.edu}

\begin{document}
\ninept 
\maketitle
\begin{abstract}
 High-fidelity speech can be synthesized by end-to-end text-to-speech models in recent years. However, accessing and controlling speech attributes such as speaker identity, prosody, and emotion in a text-to-speech system remains a challenge. This paper presents a system involving feedback constraints for multispeaker speech synthesis. We manage to enhance the knowledge transfer from the speaker verification to the speech synthesis by engaging the speaker verification network. The constraint is taken by an added loss related to the speaker identity, which is centralized to improve the speaker similarity between the synthesized speech and its natural reference audio. The model is trained and evaluated on publicly available datasets. Experimental results, including visualization on speaker embedding space, show significant improvement in terms of speaker identity cloning in the spectrogram level. In addition, synthesized samples are available online for listening. \footnote{https://caizexin.github.io/mlspk-syn-samples/index.html}
  
\end{abstract}

\section{Introduction}

Speech synthesis, also known as text-to-speech (TTS), specifies the technique that achieves the transformation from text to audio waveform. It has been widely used in our daily life, e.g., navigation systems, audiobooks, and virtual assistants. The performance of the TTS system has been further improved recently by adopting the end-to-end neural network framework \cite{mehri-srnn-2017-iclr, shen2018natural, ping2018clarinet, sotelo2017char2wav}. The end-to-end principle is applied in the TTS model by a cohesive and autoregressive chain of neural network structures that are connected by well-defined input-output features. For instance, the state-of-the-art system Tacotron2 \cite{shen2018natural} consists of an encoder-decoder architecture and a neural vocoder Wavenet\cite{oord2016wavenet}.

Extensions on these models have been developed for allowing the TTS system to control the speech characteristics. These extensions are able to enrich the expressiveness of the synthesized voice and further enhance the robustness of TTS systems. For example, Yuxuan Wang et al. proposed the style tokens to uncover the latent space regarding speech attributes that are hard to define and label \cite{wang2018style, wang2017uncovering}. The models are jointly trained with the Tacotron-based TTS architecture in an unsupervised manner. On the other hand, controlling speech attributes that have easily found labels (e.g., language, emotion, and speaker identity) have also been investigated \cite{zhang2019learning, yu2019durian, jia2018transfer}. Typically, the speech attribute is controlled with a TTS model by conditioning the synthesizer with the vector representation called embedding. 

The multispeaker TTS system is one of the extensions, which is developed to clone and manage distinct voices either seen or not seen during training. Most systems use the speaker embedding to characterize the expected voice and speaking style in the multispeaker TTS system \cite{jia2018transfer, Ji2019Multi, alvarez2019problem}, while speaker adaptation can also be used for speaker transfer TTS modeling \cite{fan2015multi}. Voice cloning by speaker adaptation acquires more data and computational resource for the target voice and usually is less robust compared with cloning by speaker embedding \cite{arik2018neural}.  The speaker verification system plays an essential role in the multispeaker TTS system for cloning unseen voices. Eliya Nachmani et al. has proposed an approach where the speaker verification system and the synthesizer are jointly trained \cite{nachmani2018fitting}. However, the knowledge for discriminative speaker representations is limited by the training dataset in this case. Then Ye Jia et al. further investigated the knowledge transfer in terms of speaker characteristics by decoupling these two tasks, where the speaker verification network is trained with a dataset that contains a larger amount of speakers but is not suitable for TTS training \cite{jia2018transfer}. The discriminative speaker embedding extracted from the speaker verification network is used for conditioning the TTS synthesizer and leads to better performance on open-set voice cloning.

Although the model proposed in \cite{jia2018transfer} increases the robustness of the synthesizer for open-set multispeaker synthesis, the speaker's similarity is not close between the synthesized speech and the speaker's natural speech. Concerning the same speaker, the embeddings extracted from synthesized speech and those extracted from natural speech may have two distinct distributions. To further transfer the knowledge from a speaker verification model to the speech synthesizer, we propose a multispeaker TTS model with the feedback constraint toward the speaker embedding space. Specifically, an added score associated with the speaker similarity is performed by the verification network for forcing the synthesizer to derive the knowledge for speaker identity cloning. The proposed method is evaluated on publicly available datasets. As demonstrated in the visualization of the embedding space, speaker embeddings extracted from our synthesized speech lies in the same cluster as those from natural speech. Therefore, the model may be useful for data augmentation and the white-box spoofing attack toward speaker verification in the future. 

This paper is organized as follows: section \ref{related_work} describes the related works in terms of speaker verification and speech synthesis. Our proposed system is presented in section \ref{system_describtion}. Experimental setup and results are shown in section \ref{result}. Finally, we conclude the paper in section \ref{conclusion}.      

\section{Related works}
\label{related_work}

\begin{figure*}[ht]
  \centering
  \includegraphics[scale=0.78]{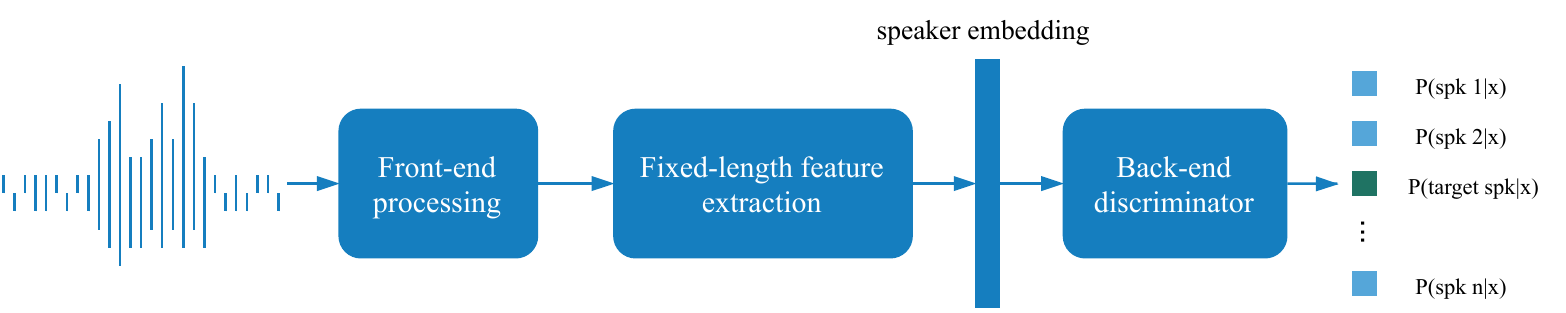}
  \caption{The overall training framework of deep speaker verification systems}
  \label{fig:spk_framework}
\end{figure*}

\subsection{Speaker verification}
Open-set multispeaker TTS highly depends on speaker representations for conditioning the synthesizer to copy the desired voice. To that end, speaker verification systems, especially text-independent systems, are often used for feature extraction regarding their discriminative speaker representations. 

In the speaker verification field, deep speaker feature learning systems proposed in recent years have achieved comparable performance or even surpassed the classical i-vector systems \cite{cai2018analysis, snyder2018x, Cai_2018_Odyssey}. The overall training architecture of the deep speaker verification system is shown in figure \ref{fig:spk_framework}. Specifically, the speaker verification model takes variable-length audio signal $x = [x_1, x_2, x_3 ... x_n]$ as input and convert the signal into frequency-domain acoustic features, e.g., filter-bank energy or Mel frequency cepstral coefficients (MFCCs). Acoustic features are then fed into a neural network-based extractor to obtain the fixed dimensional speaker representation $z \in \mathcal{R}^d $, where $d$ is the dimension of the speaker embedding. Note that the back-end discriminator here in the training phase is different from the one in the evaluation phase. The discriminator in the training phase is to classify embeddings according to their target speaker labels in order to train a discriminative speaker embedding space, while the one used in the evaluation phase is to verify if two embeddings come from the same speaker. 

Among deep speaker embedding systems that are developed with various DNN architectures \cite{snyder2018x, snyder2016deep, cai2020fly}, we follow the ResNet-based verification system \cite{cai2020fly} in our work for embedding extraction to extract time-invariant speaker embedding. 

\begin{figure*}[ht]
  \centering
  \includegraphics[scale=0.72]{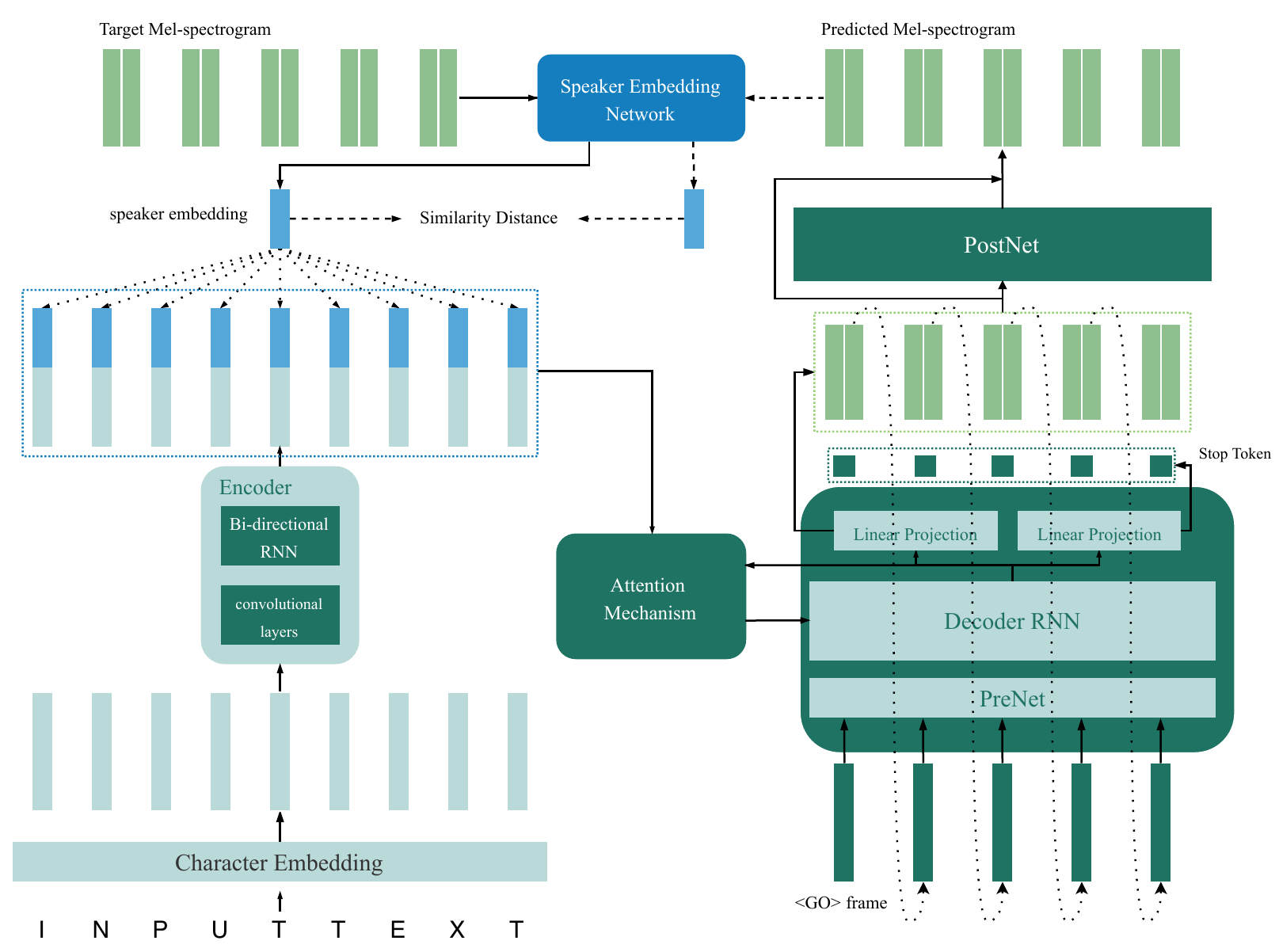}
  \caption{Proposed multi-speaker speech synthesis model}
  \label{fig:tts}
\end{figure*}

\subsection{Multispeaker speech synthesis}
Cloning and controlling speech attributes have been studied for decades in the text-to-speech (TTS) field. For voice synthesis, Yamagishi et al. proposed feature-space adaptive training for speaker-adaptive TTS \cite{yamagishi2009robust}. The system aims to reduce the size of data and the cost for building different voices when developing statistical parametric speech systems based on Hidden Markov models (HMMs). After Tacotron2 demonstrated its ability to synthesize high-quality speech that can be as natural as human speech, extensions of Tacotron2 were proposed for speech attribute cloning by conditioning the linguistic encoder output with attribute embeddings. For instance, Yuxuan Wang et al. introduced global style tokens (GSTs) as the attribute embeddings to achieve
style-control TTS synthesis \cite{wang2018style}. The proposed model in \cite{wang2018style} , where GSTs are trained in an unsupervised manner, also helps improve the speech intelligibility when it is used for multispeaker TTS training. 

On the other hand, in order to achieve zero-shot voice cloning, the speaker representation is commonly extracted by a separated model and used as the conditioned feature in multispeaker TTS models \cite{jia2018transfer, nachmani2018fitting}.  In this case, the multispeaker TTS model developed for zero-shot voice cloning consists of two models, one for speaker embedding extraction and the other for TTS conversion. When the two models are trained jointly, the TTS system yields moderate performance in synthesizing voices that are unseen in the training data \cite{nachmani2018fitting}. The reason might be because the datasets collected for speech synthesis have limited speakers, and the datasets collected for speaker analysis have no transcriptions for TTS training. Jia Ye et al. chose to train the two models individually, where the TTS model learns the speaker representation knowledge by the embedding extracted from the speaker verification model \cite{jia2018transfer}. This method improves the robustness with the ability to clone unseen voices. However, two distinct clusters, representing synthesized speech and natural speech from the same speaker, are observed in the embedding space as shown in \cite{jia2018transfer}. To further investigate this problem and enhance the knowledge transfer, we propose a model with a feedback constraint that engages the speaker embedding extractor. We show that by showing that embeddings from different speakers result in distinct distributions in the vector space, while embeddings from the same speaker, whether synthesized or natural, lie in the same cluster.

In our work, we use a speaker embedding extractor that is different from the model described in  \cite{jia2018transfer}. By the time we finalized our work, Erica et al. published a study investigating how different speaker embedding networks affects the multispeaker synthesis system \cite{cooper2019zero}. In that study, the author claim that LDE-based embedding could improve speaker similarity and naturalness. Our model has a similar speech encoder as the learnable dictionary encoding-based (LDE-based) systems described in \cite{cooper2019zero}.

\section{Methods}
\label{system_describtion}

Our proposed multispeaker TTS framework is shown in figure \ref{fig:tts}. We follow the baseline end-to-end speaker verification system presented in \cite{cai2020fly} as our embedding extraction network. The Mel-spectrogram is used as the acoustic feature for both the speaker embedding extraction system and the multispeaker TTS system. As for the speaker embedding network, ResNet34 architecture is used as the encoder network, followed by a pooling layer that calculates the mean and the standard deviation of encoder outputs. Then the speaker embedding is obtained by concatenating the mean and the standard deviation. While in the training phase, a back-end classification network consisting of two fully connected layers maps embeddings to target speakers. 

We use the tacotron-based model as the Mel-spectrogram prediction model.  The input character sequence is converted into a vector sequence by a trainable lookup table. Then the encoder, which consists of 5 convolutional layers and a bi-directional long short-term memory (BLSTM) layer, consumes the embedding sequence and delivers the memory that represents the context and linguistic characteristics of the input text. Speaker embedding, extracted from the target audio signal, is then concatenated with the encoder output memory globally as the final encoding states. 

The decoder takes steps to predict Mel-spectrograms with three modules, which are the attention mechanism, the RNN decoder and the PostNet. The attention mechanism provides the context vector for the decoder RNN to generate spectrograms that associate with specific encoder states for each decoding time step. In addition, it provides soft alignment between the input encoder states and the target Mel-spectrogram. For each decoding step, the decoder RNN predicts the spectrogram with respect to the context vector and the predicted result from the previous time step, where the previous predicted frame is taken by the PreNet module. Two linear projection layers are followed by the decoder RNN for predicting  Mel-spectrograms and stop tokens, respectively. Stop tokens are the binary sequence that specifies the valid decoding frames, where $0$ denotes a valid frame, and $1$ indicates the end of the decoding process. The PostNet takes the predicted Mel-spectrogram as input to obtain the residual parameters that are related to future context since the decoder RNN is unable to foresee future frames. The predicted Mel-spectrograms are finetuned with the PostNet, which leads to high-quality outputs.

The speaker embedding network is engaged after the PostNet during the training phase. It is added as a feedback constraint to force the TTS model to learn the speaker variety knowledge sufficiently so that the speaker characteristics extracted from synthesized Mel-spectrogram lays in the same distribution as those extracted from the natural speech from the same person. In this case, the parameters of the speaker embedding network are not updated during the training phase.

We use the cosine distance between the ground truth speaker embedding and the one extracted from the predicted Mel-spectrogram by speaker embedding network as one of the loss functions for optimizing the TTS network. Other than that, mean square error (MSE) between predicted Mel-spectrograms and the ground truth spectrogram, classification loss of the stop tokens, and the regularization loss for encoder-decoder parameters are applied to ensure correct predictions.

The Mel-spectrogram is converted back to the audio signal by the neural vocoder WaveRNN \cite{kalchbrenner2018efficient}, which is able to generate high-quality speech at fast speed.

\section{Experiments}
\label{result}

We used four publicly available datasets for training and evaluation. All data from Voxceleb1 \cite{nagrani2017voxceleb} and Voxceleb2 \cite{chung2018voxceleb2}, with more than 7, 000 speakers, are used for training the speaker verification system. The VCTK English dataset \cite{veaux2016superseded}, which contains 109 speakers with various accents, is used for TTS model training, while data from 8 speakers are randomly excluded as the VCTK test set. For each speaker in the training set, 8 utterances are randomly picked out as the VCTK validation set. 7 speakers from the Librispeech dataset \cite{panayotov2015librispeech} are randomly selected as the cross-domain test set. All audios are downsampled to 16 kHz in our experiments.

\begin{figure}[h]
  \centering
  \includegraphics[width=0.47\textwidth]{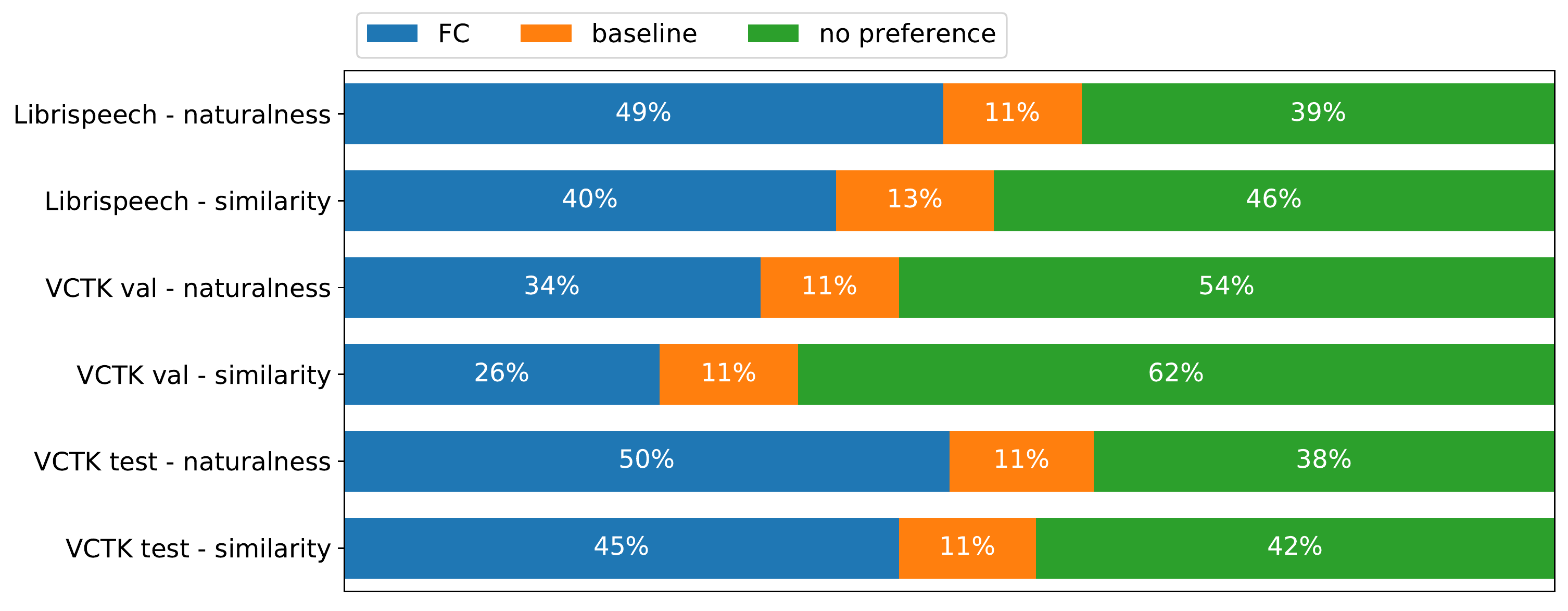}
  \caption{Subjective preference result}
  \label{fig:objeva}
\end{figure}

We evaluate the performance by comparing the proposed system, which has added feedback constraint (FC), with the multispeaker TTS baseline system without FC. The two systems are identified by \textbf{`baseline'} and \textbf{`FC'} in the following subjective and objective results. We first trained the baseline model until it can synthesize intelligible speech. Then the FC model is trained from the pre-trained baseline model while engaging the speaker embedding network. Both models are then trained to the same total training steps with the same batch size. 

\begin{figure*}[ht]
  \includegraphics[scale=0.27]{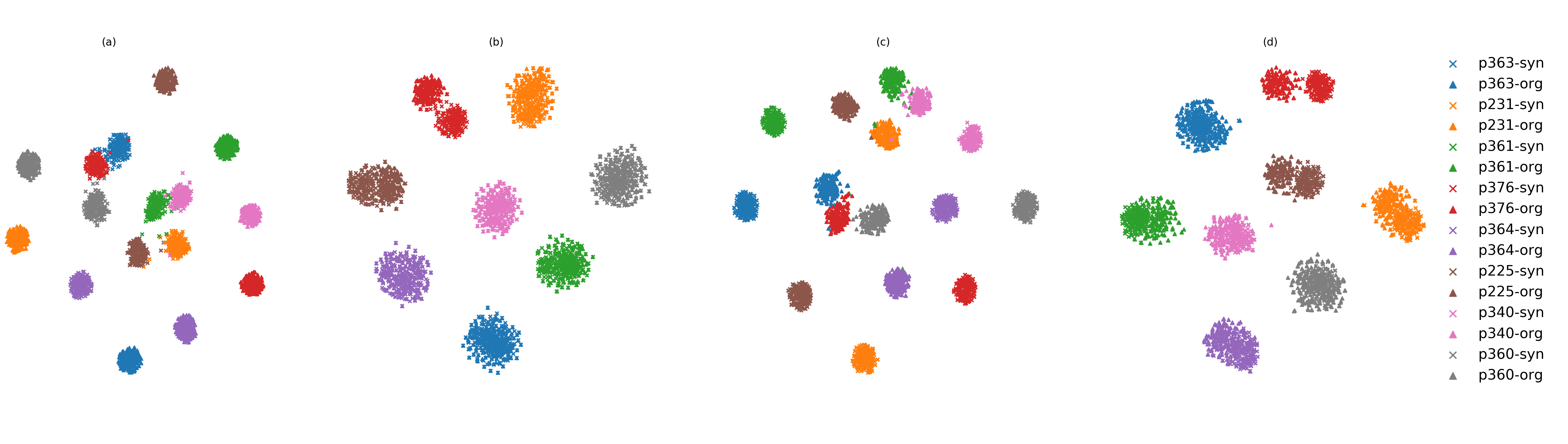}
  \caption{Speaker embedding visualization by t-SNE for the VCTK test set. (a) baseline \& text-dependent speaker embedding; (b) FC \& text-dependent speaker embedding; (c) baseline \& text-independent speaker embedding; (d) FC \& text-independent speaker embedding; }
  \label{fig:vctk_textdep}
\end{figure*}

\subsection{Subjective evaluation}
We asked 12 people to choose their preferable speech for pairs that contain speech synthesized by both systems. Audios in each pair are synthesized with the same text content and the conditioned embedding from the same reference audio. Each person chose their preference concerning speaker similarity and naturalness from 38 pairs, which are randomly selected from the VCTK test set, the VCTK validation set, and the Librispeech set. Preference results are shown in figure \ref{fig:objeva}. For subjective evaluation, the FC system outperforms the baseline system on all three evaluation sets. For seen speakers in the training data, the speech synthesized by both systems is close. Hence people do not have a preferred choice for more than $50\%$ pairs in the VCTK validation set. Given these points, both systems can copy seen voices well, while the FC system obtains better performance on unseen voice cloning.

\begin{table}[]
\caption{Objective evaluation results}
\label{tbl:objeva}
\begin{tabular}{|l|c|c|c|}
\hline
                                                 & Systems  & \begin{tabular}[c]{@{}c@{}}SV-EER (\%) \\ Dep / Indep \end{tabular}  & \begin{tabular}[c]{@{}c@{}}Average \\ cosine similarity \\ Dep\ /\ Indep \end{tabular} \\ \hline
\multicolumn{1}{|c|}{\multirow{3}{*}{VCTK test}} & natural  & 1.76     & -                   \\ 
\multicolumn{1}{|c|}{}                           & baseline & 14.72 / 13.18     & 0.403 / 0.333      \\ 
\multicolumn{1}{|c|}{}                           & FC       & 8.22 / 7.68     & 0.764 / 0.577             \\ \hline
\multirow{3}{*}{VCTK val}                        & natural  & 1.61     & -                   \\ 
                                                 & baseline & 9.22 / 8.23     & 0.472 / 0.394  \\ 
                                                 & FC       & 5.02 / 3.42     & 0.842 / 0.67    \\ \hline
\multirow{3}{*}{Librispeech}                     & natural  & 5.26     & -          \\ 
                                                 & baseline & 26.84 / 26.46     & 0.222 / 0.139        \\ 
                                                 & FC       & 16.54 / 16.11     & 0.626 / 0.389     \\ \hline
\end{tabular}
\end{table}

\subsection{Objective evaluation}
For each utterance from all evaluation datasets, we synthesized speech according to the given transcript and the embedding extracted from the original speaker's voice. Two different synthesized results were collected for each utterance. Although both are synthesized with the same reference voice, one is synthesized based on the speaker embedding extracted from the utterance that has the exact same content, while the other result is synthesized with the speaker embedding extracted from a randomly selected utterance with different content. These are identified as text-dependent \textbf{(Dep)} result and text-independent \textbf{(Indep)} result in table \ref{tbl:objeva}. Speaker verification equal error rate \textbf{(SV-EER)} is used to evaluate the speaker discrimination performance for a set of embeddings. We randomly generate enrollment-verification pairs for each experiment, where half of the trials are cross-speaker pairs. We also compute the average cosine similarity between embeddings extracted from synthesized speech and the ground truth embeddings to measure the speaker similarity performance objectively. As shown in table \ref{tbl:objeva}, the FC system obtains significantly lower EERs than the baseline system on all evaluation sets, whether text-dependent or text-independent. The FC system also has higher average cosine similarities than the baseline system. In either case, we can conclude that the voice synthesized by the FC system is more close to the reference voice than the baseline system. The similarity is improved with the feedback constraint network. 

Likewise, we can visualize the results from the embedding space by the t-Distributed Stochastic Neighbor Embedding (t-SNE), as shown in figure \ref{fig:vctk_textdep}. The utterances synthesized by the baseline system with reference embeddings from the same speaker is in the same cluster, but do not have the same distribution with reference embeddings, even is closer to another speaker. For example, as shown in figure \ref{fig:vctk_textdep} $(a)$, embeddings from `p225-syn' have a distribution that is close to `p376-org' other than its reference speaker `p225-org'. For the FC system, the embeddings extracted from the same voice, either synthesized or natural, lie in the same distribution in the embedding space. Therefore, the synthesized voice is more close to the original speaker for utterances synthesized by the FC system.

    
  

\section{Conclusion}
\label{conclusion}
In this paper, a multispeaker TTS approach that explores the use of a speaker verification system is presented. A trained speaker verification system is incorporated into the TTS framework acting as the feedback constraint to facilitate voice cloning. Experimental evaluations, including both subjective and objective evaluations, demonstrate that our proposed system enhances the knowledge transfer from speaker verification to speech synthesis.  Accordingly, our proposed method achieves significant improvement regarding voice cloning, which can be used for data augmentation or white-box spoofing attack in the future. 

\footnotesize{$\ $}

\footnotesize{\noindent\textbf{Acknowledgments} This research is funded in part by Duke Kunshan University.\par}

\vfill\pagebreak

\bibliographystyle{IEEEtran}

\bibliography{mybib}


\end{document}